\documentclass[letterpaper]{article} % DO NOT CHANGE THIS
\usepackage{aaai2026}
\usepackage{times}  % DO NOT CHANGE THIS
\usepackage{helvet}  % DO NOT CHANGE THIS
\usepackage{courier}  % DO NOT CHANGE THIS
\usepackage[hyphens]{url}  % DO NOT CHANGE THIS
\usepackage{graphicx} % DO NOT CHANGE THIS
\usepackage{natbib}  % DO NOT CHANGE THIS AND DO NOT ADD ANY OPTIONS TO IT
\usepackage{caption} % DO NOT CHANGE THIS AND DO NOT ADD ANY OPTIONS TO IT
\usepackage{algorithm}
\usepackage{algorithmic}
\usepackage{makecell}   % 用于 \Xhline，可以控制粗细的 \hline
\usepackage{amsmath}
\usepackage{newfloat}
\usepackage{listings}
\usepackage{booktabs}
\usepackage{caption}

\DeclareCaptionStyle{ruled}{labelfont=normalfont,labelsep=colon,strut=off} % DO NOT CHANGE THIS
\floatstyle{ruled}
\newfloat{listing}{tb}{lst}{}
\floatname{listing}{Listing}
\usepackage[labeled]{multibib}
\newcites{A}{Appendix References}

\usepackage{multirow}

\title{Plug-and-Play Clarifier: A Zero-Shot Multimodal Framework for \\Egocentric Intent Disambiguation}
\author{
Sicheng Yang\textsuperscript{\rm 1}, Yukai Huang\textsuperscript{\rm 2}, Weitong Cai\textsuperscript{\rm 3}, Shitong Sun\textsuperscript{\rm 3}, 
You He\textsuperscript{\rm 1 *},%\thanks{Corresponding author: heyou@mail.tsinghua.edu.cn},
\\Jiankang Deng\textsuperscript{\rm 4}, Hang Zhang\textsuperscript{\rm 2}, Jifei Song\textsuperscript{\rm 5}, Zhensong Zhang\textsuperscript{\rm 2} 
}
\affiliations{
\textsuperscript{\rm 1}Shenzhen International Graduate School, Tsinghua University
\textsuperscript{\rm 2}Independent Researcher\\
\textsuperscript{\rm 3}Queen Mary University of London
\textsuperscript{\rm 4}Imperial College London
\textsuperscript{\rm 5}University of Surrey

yangsc25@mails.tsinghua.edu.cn, 
heyou@mail.tsinghua.edu.cn\\
\textsuperscript{\rm *}Corresponding author
}

\usepackage{bibentry}
\begin{document}

\maketitle

\begin{abstract}
The performance of egocentric AI agents is fundamentally limited by multimodal intent ambiguity. This challenge arises from a combination of underspecified language, imperfect visual data, and deictic gestures, which frequently leads to task failure. Existing monolithic Vision-Language Models (VLMs) struggle to resolve these multimodal ambiguous inputs, often failing silently or hallucinating responses. To address these ambiguities, we introduce the \textbf{Plug-and-Play Clarifier}, a zero-shot and modular framework that decomposes the problem into discrete, solvable sub-tasks. Specifically, our framework consists of three synergistic modules: (1) a text clarifier that uses dialogue-driven reasoning to interactively disambiguate linguistic intent, (2) a vision clarifier that delivers real-time guidance feedback, instructing users to adjust their positioning for improved capture quality, and (3) a cross-modal clarifier with grounding mechanism that robustly interprets 3D pointing gestures and identifies the specific objects users are pointing to. Extensive experiments demonstrate that our framework improves the intent clarification performance of small language models (4--8B) by approximately 30\%, making them competitive with significantly larger counterparts. We also observe consistent gains when applying our framework to these larger models. Furthermore, our vision clarifier increases corrective guidance accuracy by over 20\%, and our cross-modal clarifier improves semantic answer accuracy for referential grounding by 5\%. Overall, our method provides a plug-and-play framework that effectively resolves multimodal ambiguity and significantly enhances user experience in egocentric interaction.

\end{abstract}

% Uncomment the following to link to your code, datasets, an extended version or similar.
% You must keep this block between (not within) the abstract and the main body of the paper.
% \begin{links}
    % \link{Code}{https://aaai.org/example/code}
    % \link{Datasets}{https://aaai.org/example/datasets}
    % \link{Extended version}{https://aaai.org/example/extended-version}
% \end{links}

% 7 pages
\section{Introduction}  % 0.25 - 1.5 / 1.75 / 2 pages

\begin{figure}[!t]
    \centering
    \includegraphics[width=0.999\linewidth]{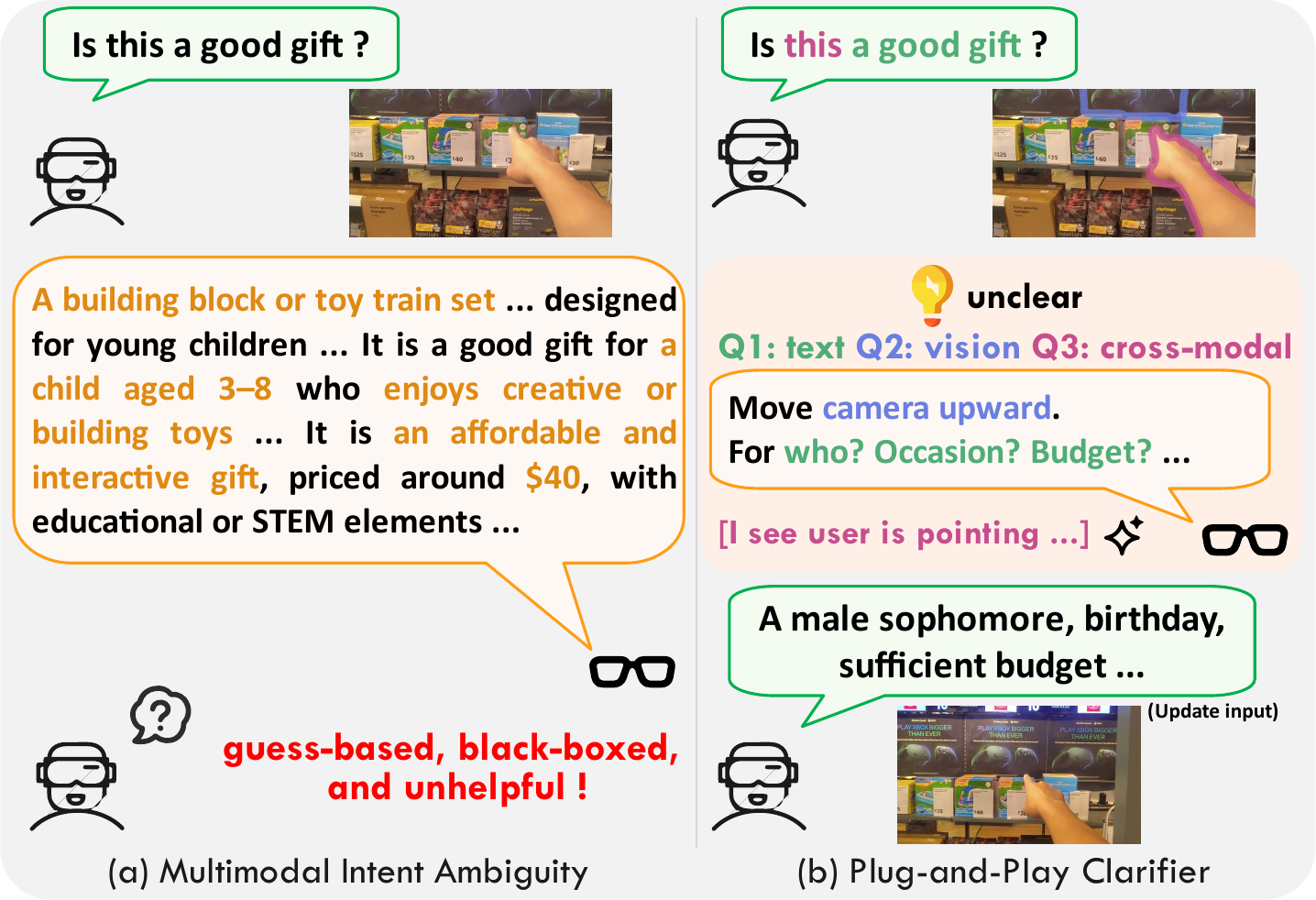}
    \caption{
    Our Clarifier framework resolves the multimodal ambiguous query, ``Is this a good gift?''
(a) Multimodal Intent Ambiguity: A standard AI defaults to a guess, making assumptions about the recipient (e.g., a child), their interests, and the user's budget. This ``black box'' approach is unhelpful because if the assumptions are wrong, the recommendation is useless.
(b) Plug-and-Play Clarifier: 
Our system avoids guessing. It first identifies that key information (recipient, occasion, budget), missing visual context, and pointing gestures between modalities. It then proactively asks clarifying questions and provides camera feedback ("For who? Move camera upward..."). Once the user provides the necessary context, the system can deliver a relevant and genuinely helpful recommendation.}
    \label{fig:enter-label}
\end{figure}

Egocentric AI agents, particularly those embedded in wearable devices such as AI glasses, are emerging as a new and significant area of human-computer interaction.
The goal is to develop an always-on cognitive partner that perceives the world from the user's first-person perspective. 
Such an agent could understand user goals and provide seamless assistance for daily physical tasks, including assembling furniture, cooking complex recipes, or navigating unfamiliar environments.
By operating from this viewpoint, these agents can deliver contextually-rich, proactive support that is tightly integrated with the user's activities \cite{DBLP:journals/corr/abs-2503-15275}.

However, this vision faces a fundamental challenge: \textbf{multimodal intent ambiguity}. Unlike interactions with text-based chatbots, egocentric interaction is inherently noisy, fast-paced, and underspecified. A simple spoken command, such as ``What about that one?'', is inherently ambiguous. Which object is ``that one''? Is the object of interest clearly visible, or is it blurry or partially occluded in the camera feed? This ambiguity stems from a combination of sources: underspecified natural language, imperfect visual data from the wearable camera, and deictic gestures like pointing \cite{DBLP:conf/cvpr/HuangLZJ16}. As a result, even state-of-the-art AI assistants frequently misinterpret user intent, leading to frustrating and unproductive task failures \cite{DBLP:journals/corr/abs-2404-16244, weidemann2021role}.

Current approaches, dominated by monolithic end-to-end Vision-Language Models (VLMs), fall short in robustly addressing this challenge \cite{DBLP:conf/acl/HuangQQLJXW25}. 
While large models like GPT-4o demonstrate strong general multimodal understanding, they function as opaque ``black boxes.'' 
When presented with ambiguous, mixed-signal inputs, they tend to ``hallucinate'' an interpretation or fail silently, as they lack a mechanism to proactively seek clarification. 
Requiring a single, massive model to concurrently manage linguistic interpretation, spatial reasoning, and visual quality assessment is inherently brittle. 
This approach is also computationally expensive, making it a poor fit for resource-constrained wearable devices.

To overcome these limitations, we propose a shift from monolithic reasoning to a structured, modular, and interactive approach. We introduce the \textbf{Plug-and-Play Clarifier}, a zero-shot, multimodal framework that resolves ambiguity in egocentric interactions. Instead of relying on a single model's opaque reasoning, our framework decomposes the problem into discrete sub-tasks managed by an explicit, programmatic control loop.
This architecture integrates three core modules: 
(1) a \textit{text clarifier} with dialogue-driven reasoning that clarifies user intent through a structured, step-by-step conversation; 
(2) a \textit{vision clarifier} that assesses visual input quality (e.g., framing, clarity) and provides real-time corrective feedback; and 
(3) a \textit{cross-modal clarifier} with grounding mechanism that interprets 3D pointing gestures by casting a geometric ray into the scene to localize the referenced object. 
Our framework enhances existing foundation models without requiring fine-tuning. This modular approach proves more robust and efficient than end-to-end black-box models for complex real-world tasks.
Our code and demos are available online\footnote{\url{https://github.com/YoungSeng/plug-and-play-clarifier}}.
We summarize our contributions as follows:
% \begin{itemize}
%     \item We propose a zero-shot and plug-and-play framework that resolves multimodal intent ambiguity in egocentric interaction through problem decomposition and interactive clarification.
%     \item We show that our framework improves the intent clarification accuracy of small language models (4--8B) by ~30\% on textual tasks, making them competitive with much larger models.
%     \item On our newly introduced VRA-Ego benchmark, we demonstrate that our individual modules achieve significant improvements: the vision clarifier increases corrective guidance accuracy by over 20\%, and the cross-modal clarifier with 3D pointing module improves semantic grounding accuracy by 5\%, outperforming strong monolithic baselines.
%     % We demonstrate that our individual modules achieve significant improvements: the vision clarifier increases corrective guidance accuracy by over 20\%, and the cross-modal clarifier with 3D pointing module improves semantic grounding accuracy by 5\%, outperforming strong monolithic baselines.
%     \item We validate that a hybrid architecture combining the generative capabilities of LLMs with algorithmic modules is a more robust, efficient, and interpretable approach for building reliable egocentric AI.
% \end{itemize}
(1) We propose a zero-shot and plug-and-play framework that resolves multimodal intent ambiguity in egocentric interaction through problem decomposition and interactive clarification.
(2) We show that our framework improves the intent clarification accuracy of small language models (4--8B) by ~30\% on textual tasks, making them competitive with much larger models.
(3) On our newly introduced VRA-Ego benchmark, we demonstrate that our individual modules achieve significant improvements: the vision clarifier increases corrective guidance accuracy by over 20\%, and the cross-modal clarifier with 3D pointing module improves semantic grounding accuracy by 5\%, outperforming strong monolithic baselines.
(4) We validate that a hybrid architecture combining the generative capabilities of LLMs with algorithmic modules is a more robust, efficient, and interpretable approach for building reliable egocentric AI.

% Our work validates that for complex, real-world tasks, a modular and transparent reasoning process is a more robust and efficient strategy than end-to-end black-box reasoning, paving the way for more capable and reliable egocentric AI assistants.

\section{Related Work}  % 2 - 2.5 pages
\subsection{Intent Clarification in Dialogue Systems}

The field of intent clarification has moved from traditional approaches, such as programmatic slot-filling and templated questions~\cite{DBLP:conf/chi/ZhangAHB25, DBLP:journals/corr/abs-2503-01940, li5168033mdsd}, towards end-to-end systems that use Large Language Models (LLMs)~\cite{DBLP:conf/acl/ZhangQDHLLJLC24, DBLP:conf/acl/QianHZDQCZZL0024}. 
To address the inconsistent performance of basic prompting methods like Chain-of-Thought (CoT), subsequent work has focused on incorporating more structure. 
For example, some methods require the LLM to first determine the type of ambiguity before generating a response~\cite{DBLP:conf/sigir/TangSG25}. 
Others involve developing system-level frameworks that use specialized classifiers~\cite{DBLP:journals/corr/abs-2502-00537} or Bayesian inference~\cite{DBLP:journals/corr/abs-2504-09665} to guide the clarification process.
Recent benchmarks~\cite{DBLP:journals/corr/abs-2502-01523} and datasets~\cite{DBLP:conf/emnlp/AliannejadiKC0B21} support a move towards more structured and interpretable disambiguation, yielding specialized frameworks for enterprise~\cite{DBLP:conf/aaai/MurzakuLTMC025} and knowledge-graph applications.

Despite these advances, existing work largely falls into two categories. 
The first is end-to-end LLM solutions, which can lack transparency and rely heavily on the model's own reasoning capabilities \cite{DBLP:journals/tmlr/NguyenBSKNL25, de2025unveiling}.
The second is applications designed for text-only scenarios, such as travel planning~\cite{DBLP:conf/emnlp/Zhang0RNC24, DBLP:journals/corr/abs-2505-11533} or e-commerce~\cite{DBLP:conf/wsdm/DammuAP25}.
In contrast, we introduce a hybrid approach that uses a programmed, zero-shot external control loop to guide the clarification process. 
This architecture provides transparency and control by design, while reducing the reasoning load placed on the LLM. 
Furthermore, we are the first to apply such a framework to the challenging domain of multimodal, first-person interactions in the physical world, extending intent clarification beyond traditional text-based interfaces.

\subsection{Egocentric Vision and Interaction}

Egocentric Vision (EGV) provides a first-person perspective to infer user intent from tasks such as action recognition~\cite{DBLP:conf/wacv/ShiotaTKSA24}, hand-object interaction~\cite{DBLP:conf/iccv/Xu0HLLTT23, DBLP:conf/iclr/XuWDSZJ25}, and scene understanding~\cite{DBLP:journals/corr/abs-2503-15275}. 
This has led to large-scale projects like EgoLife and EgoM2P, which aim to build assistive agents from complex, real-world multimodal data~\cite{DBLP:journals/corr/abs-2506-09995}.
However, a key challenge in EGV is that its data is inherently noisy and ambiguous, unlike the curated data used in standard VQA or VLM benchmarks~\cite{DBLP:conf/iccvw/Fan19, DBLP:conf/cvpr/PerrettDSEPPLGB25}.
Even state-of-the-art models like Gemini Pro struggle with it, revealing the limitations of current VLMs when processing uncurated, real-world egocentric video \cite{google2025gemini25pro}.

Pointing gestures are a critical signal in Egocentric Vision (EGV)~\cite{DBLP:conf/icra/Das21}. 
While modern systems integrate gestures with language for 3D scene understanding~\cite{DBLP:conf/cvpr/ManeWSSSM25}, many earlier approaches were limited. 
For instance, some work relied only on the 2D screen position of a fingertip~\cite{DBLP:conf/smc/HuangLJZ15, DBLP:conf/cvpr/HuangLZJ16}. 
This simplification ignores the 3D pointing vector, which is essential for determining what a person is referring to. 
As a result, the ambiguity inherent in the pointing action itself remained largely unaddressed.

Most existing research passively attempts to interpret noisy and ambiguous inputs, with a recent trend of proactive agents only just beginning to emerge~\cite{DBLP:conf/iclr/LuYQCLWWCZLLWLL25}. 
In contrast, we propose a proactive interaction framework. 
Rather than analyzing potentially flawed data after the fact, our system actively identifies input ambiguity—such as an imprecise pointing gesture—and uses real-time multimodal feedback to guide the user toward providing a clearer, more reliable input. 
We argue this shift from passive processing to active guidance is a key step toward more robust and intuitive first-person interactive systems~\cite{DBLP:journals/corr/abs-2506-05904}.

\subsection{Multimodal Reasoning with Large Models}

While state-of-the-art Vision-Language Models (VLMs) like GPT-4o~\cite{DBLP:journals/corr/abs-2410-21276}, Gemini 2.5 Pro~\cite{google2025gemini25pro}, and Grok-3~\cite{xai2025grok3} perform well on general multimodal tasks~\cite{DBLP:journals/corr/abs-2502-13923}, they are known to be unreliable for tasks requiring precise spatial or geometric reasoning, often leading to hallucinations~\cite{DBLP:conf/emnlp/MouselinosMM24, DBLP:conf/siggrapha/Feng0WLW24, DBLP:conf/acl/HuangQQLJXW25, ramachandran2025well}.
One line of work addresses this by improving the monolithic models themselves, for instance, through specialized pre-training for high-resolution visuals~\cite{DBLP:journals/corr/abs-2412-19437}, incorporating fine-grained object grounding~\cite{DBLP:journals/corr/abs-2502-13923}, or using native multimodal pre-training methods~\cite{zhu2025internvl3}.
In contrast, our work follows an alternative approach that builds hybrid systems. These systems decompose problems to combine the semantic understanding of LLMs with the precision of specialized algorithms~\cite{DBLP:conf/emnlp/0001BZGVJ024, DBLP:conf/naacl/SharmaDKZP25, DBLP:journals/corr/abs-2502-03671}.

Our work differs from prior efforts by implementing a hybrid design as an interactive, iterative clarification loop. 
Unlike approaches that attempt to generate all clarification questions in a single turn, our method refines its understanding through a step-by-step dialogue to resolve ambiguity. 
This iterative process is not only more efficient but, crucially, allows smaller, resource-constrained models to handle complex multimodal tasks that would otherwise be beyond their capabilities.

\section{Methodology}  % 2.5 - 4 pages (1.5-2.5 pages)

\subsection{Text-based Intent Clarification}

We propose a zero-shot, \textbf{dialogue-driven reasoner} for resolving ambiguous text intents.
Our approach extends Chain-of-Thought (CoT) prompting \cite{DBLP:journals/corr/abs-2411-14466, DBLP:journals/corr/abs-2503-21544, DBLP:conf/ecir/LuMRAC25} by guiding a Large Language Model (LLM) to iteratively clarify a user's goal through conversation. 
This process relies entirely on in-context learning and does not require any task-specific fine-tuning \cite{DBLP:conf/chi/ZhangAHB25, DBLP:conf/ecir/LuMRAC25}.

Our method operates iteratively. In each turn $t$, given the initial user request $U_0$ and the conversation history $H_t$, the LLM first analyzes the user's intent to identify known ($K_t$) and missing ($M_t$) pieces of information:
\begin{equation}
(K_t, M_t) = \text{Analyze}(U_0, H_t)
\label{eq:analyze}
\end{equation}
To maintain an efficient dialogue, the model then assigns a priority $p(m)$ (e.g., critical, important) to each missing item $m \in M_t$ and selects the highest-priority item $m^*_t = \underset{m \in M_t}{\arg\max} \, p(m)$ to ask about next.
The LLM generates a question $Q_t$ based on $m^*_t$. 
The user's answer, $A_t$, is added to the history ($H_{t+1} = H_t \cup \{(Q_t, A_t)\}$), and the process repeats. 
The loop terminates when no high-priority information is missing from $M_t$. 
Finally, the LLM uses the complete history $H_{final}$ to generate a structured and actionable summary of the user's intent. 
The entire process is driven by structured in-context prompts, requiring no updates to the LLM's parameters \cite{DBLP:conf/nips/KojimaGRMI22}.

\begin{figure}[!t]
    \centering
    \includegraphics[width=0.995\linewidth]{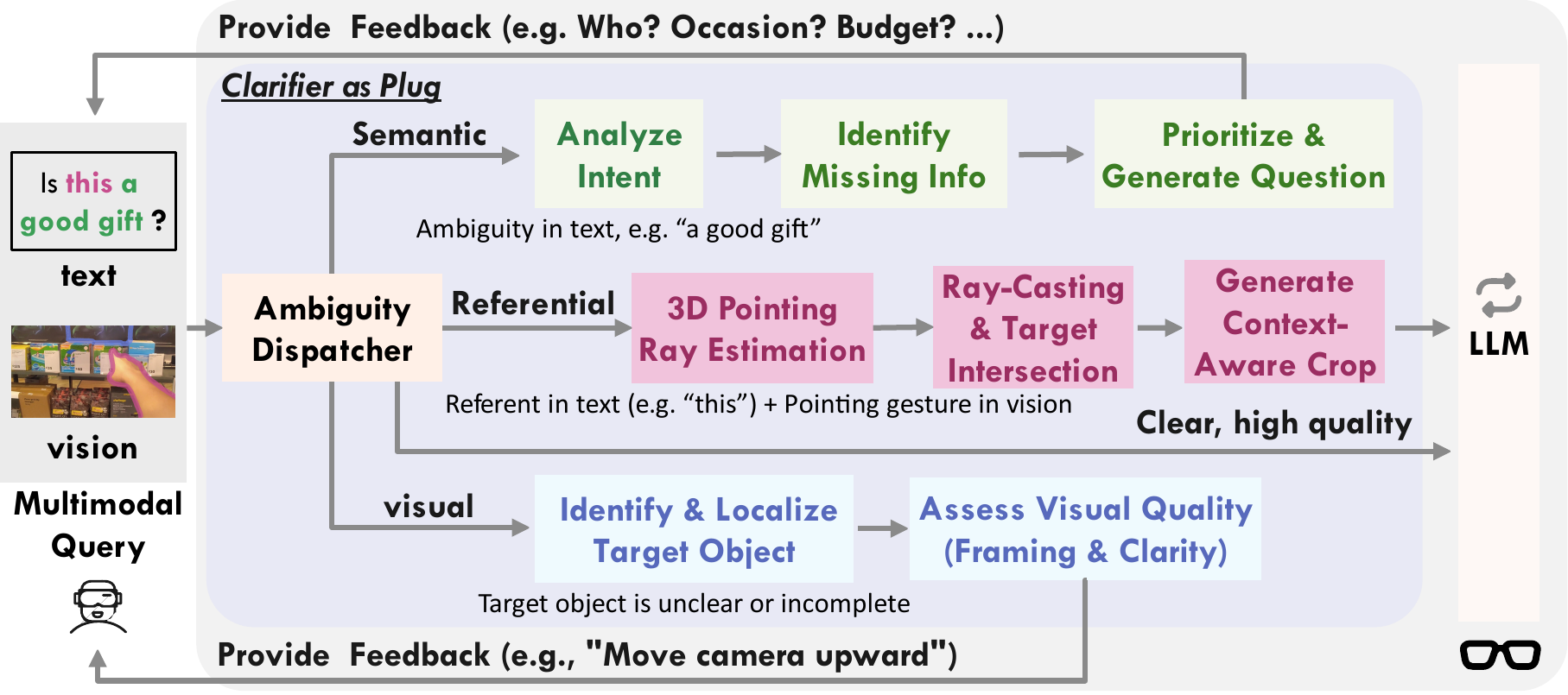}
    \caption{An overview of our clarification pipeline, a plug-in module for resolving ambiguous multimodal queries. The pipeline identifies and addresses three types of underspecification: (1) semantic ambiguity in language (e.g., ``a good gift'') is clarified through dialogue; (2) visual ambiguity from unclear object views is handled by requesting a better view; and (3) referential ambiguity from pointing gestures (e.g., ``this'') is improved by adaptive image cropping.}
    \label{fig:three - method - pipeline}
\end{figure}

\begin{figure*}[!ht]
    \centering
    \includegraphics[width=0.99\linewidth]{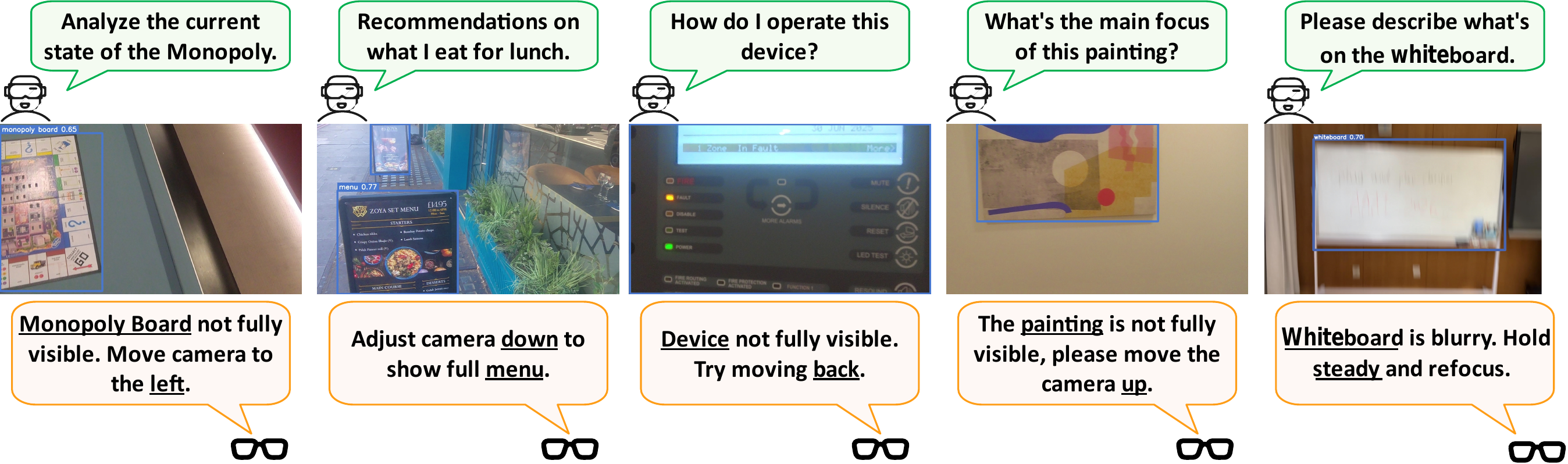}
    \caption{Overview of our vision-based clarification module. Given a user's query about a physical object, the system first identifies the target class (e.g., ``menu") using an VLM. An open-set detector then localizes the object in the image frame. Subsequently, the visual quality is assessed for framing integrity and clarity. If issues like improper framing or blurriness are detected, the system provides real-time corrective feedback to the user, ensuring high-quality visual input before proceeding.}
    \label{fig:vision_clarification_overview}
\end{figure*}

\begin{figure*}[t]
    \centering
    \includegraphics[width=0.99\linewidth]{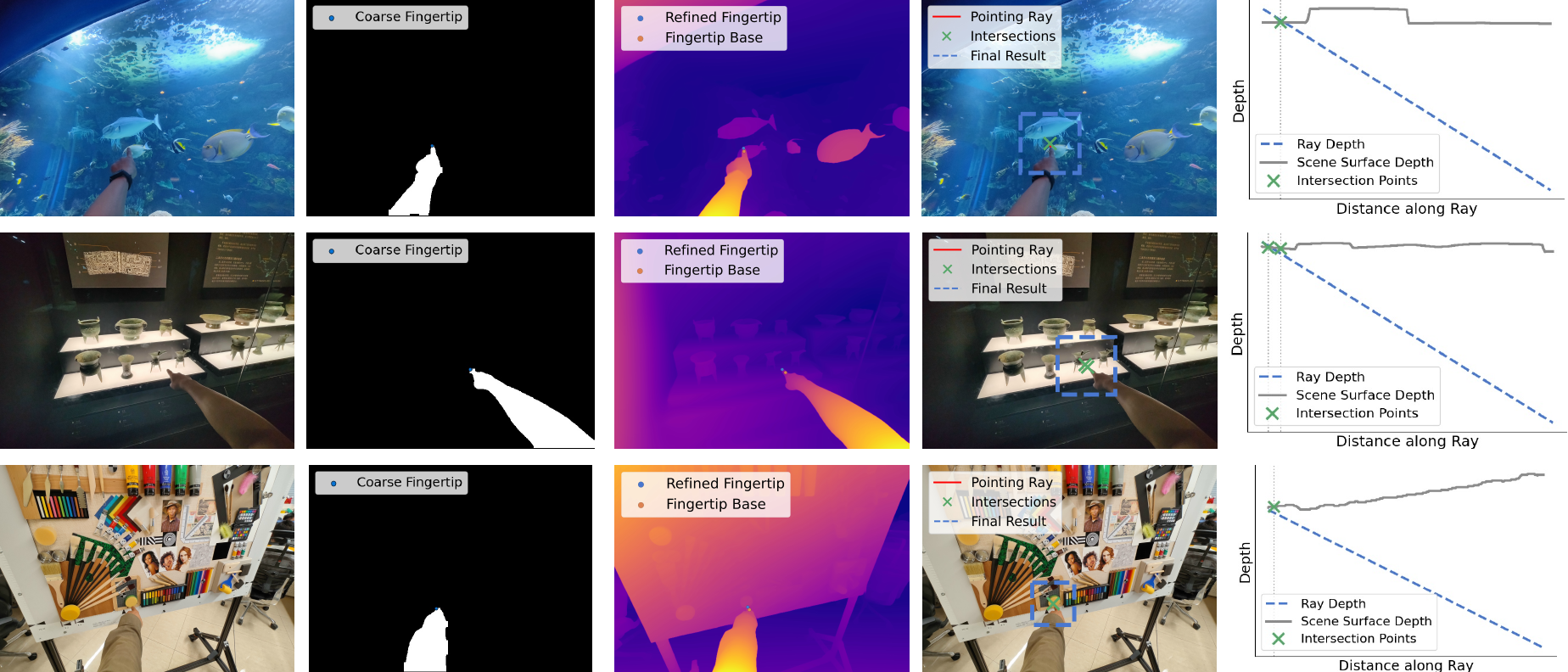}
    \caption{Our multi-stage pipeline for resolving cross-modal referential ambiguity. From a single image, we (1) estimate a 3D pointing ray from the user's hand gesture, (2) cast this ray into the scene to find a 3D intersection point, and (3) identify the target object and generate a context-aware crop containing both the hand and the object, which is then passed to a VLM for final interpretation.}
    \label{fig:pointing}
\end{figure*}

\subsection{Vision-based Intent Clarification}

To address the visual ambiguity inherent in first-person interactions, our framework proposes \textbf{a vision clarifier}.
As shown in Figure \ref{fig:vision_clarification_overview}, this module is designed to identify the user's intended visual target and verify its image quality before downstream processing.

The process begins when an VLM parses the query $U_v$ to extract the target object's class label $c$ in a zero-shot manner:     %  user's 
\begin{equation}
c = \text{ExtractEntity}(U_v)
\label{eq:extract_entity}
\end{equation}

This label $c$ guides an open-set object detector \cite{DBLP:conf/eccv/LiuZRLZYJLYSZZ24} to find the object in the image frame $I$, producing a bounding box $B$. 
If the detector fails to find the object, the system prompts the user to point the camera at the target.

Once the object is localized, the system evaluates the quality of the region of interest (ROI) inside $B$. 
First, it checks for proper \textit{framing}. 
The object's relative area must be within a predefined range $[\tau_\text{small}, \tau_\text{large}]$, and its bounding box $B$ must not be clipped by the image edges (margin $\delta_\text{edge}$). 
Second, it evaluates \textit{image clarity} using a score that combines two metrics: the variance of the Laplacian ($\mathcal{C}_\text{lap}$) to detect focus blur \cite{bansal2016blur} and the FFT high-frequency energy ratio ($\mathcal{C}_\text{fft}$) to detect motion blur \cite{shi2014discriminative}. 
The final clarity score is a weighted sum of the normalized values:
\begin{equation}
\mathcal{S}_\text{clarity}(I_B) = w_\text{lap} \cdot \text{Norm}(\mathcal{C}_\text{lap}(I_B)) + w_\text{fft} \cdot \text{Norm}(\mathcal{C}_\text{fft}(I_B))
\label{eq:clarity_score}
\end{equation}
If the framing is poor or $\mathcal{S}_\text{clarity}$ is below the threshold $\tau_\text{blur}$, the system gives the user specific, corrective feedback (e.g., ``Move further away'', ``Hold steady'').       % the clarity score
This feedback loop repeats until a well-framed, clear image is obtained, improving the reliability of subsequent vision-based tasks.

\subsection{Cross-Modal Referential Clarification}

To resolve referential ambiguity from pointing gestures (e.g., "that object"), our framework introduces a \textbf{cross-modal grounding mechanism} that converts the user's pointing direction into a 3D ray to identify specific objects in the scene.
This multi-stage pipeline, depicted in Figure \ref{fig:pointing}, ensures robust grounding from a single egocentric image.

First, we estimate the 3D pointing ray \cite{DBLP:conf/iccv/NakamuraKNN23}. From a single image $I$, we concurrently generate a hand segmentation \cite{DBLP:journals/corr/abs-2503-07465} mask $M_\text{hand}$ (via pose estimation \cite{DBLP:journals/corr/abs-2410-17725, DBLP:journals/mta/Afifi19}) and a dense depth map $D$ (via monocular depth estimation \cite{DBLP:conf/nips/YangKH0XFZ24}). The pointing vector is derived from two keypoints on the contour of $M_\text{hand}$: the fingertip ($p_\text{tip}^\text{2D}$) and base ($p_\text{base}^\text{2D}$). To improve robustness against ambiguous contours, we refine the location of $p_\text{tip}^\text{2D}$ by analyzing depth gradients along the finger's axis, ensuring it lies on the foreground finger. These 2D points are then unprojected to 3D using the depth map $D$. The final pointing ray originates at $p_\text{base}^\text{3D}$ with a normalized direction vector $\vec{v}$:
\begin{equation}
\vec{v} = \frac{p_\text{tip}^\text{3D} - p_\text{base}^\text{3D}}{\| p_\text{tip}^\text{3D} - p_\text{base}^\text{3D} \|}
\label{eq:pointing_vector}
\end{equation}

Next, we perform target localization via ray-casting. 
We cast the ray into the 3D scene to find the intersection point $P_\text{intersect}$. This point is determined by finding the point $p$ along the ray whose depth, $\text{depth}(p)$, most closely matches the value in the scene's depth map $D$ at the corresponding 2D projection of $p$, denoted as $p_{xy}$. 
This process is formulated as:
\begin{equation}
P_\text{intersect} = \underset{p \in \text{Ray}(p_\text{base}^\text{3D}, \vec{v})}{\arg\min} | \text{depth}(p) - D(p_{xy}) |
\label{eq:ray_casting}
\end{equation}
% subject to the difference being within a collision threshold $\tau_\text{collision}$.
% This process is defined as
% \begin{equation}
% P_{\text{intersect}} = \arg\min_{p \in \text{Ray}(p_{\text{base}}^{3D}, \vec{v})} \big| \text{depth}(p) - D(p_{xy}) \big|
% \label{eq:ray_casting}
% \end{equation}
subject to $\big|\text{depth}(p) - D(p_{xy})\big| \le \tau_{\text{collision}}$.

Finally, we conduct object identification and context-aware cropping. Since simple point-based grounding is often brittle in cluttered scenes, our approach generates a depth-aware Region of Interest (ROI). We define a bounding box $B_\text{target}$ centered on the 2D projection of $P_\text{intersect}$, with dimensions dynamically scaled by its depth. This adapts the ROI size to the object's distance. To resolve the user's query $U_v$ while preserving deictic context, we perform a context-aware crop: we compute a consolidated bounding box, $B_\text{context}$, that minimally encloses both the target ROI $B_\text{target}$ and the user's hand. The resulting crop $I[B_\text{context}]$ is passed with the query $U_v$ to the VLM, mitigating background noise while retaining the crucial gesture-object link. As validated by our ablation studies (see supplementary material), this method is more robust than direct grounding or simple visual overlays.

\section{Experiments}   % 4 - 6.75 pages

\subsection{Experimental Setup}

\subsubsection{Dataset.}

Our evaluation leverages established benchmarks for textual disambiguation and introduces a novel benchmark to address ambiguities unique to first-person vision. For textual tasks, we use IN3 \cite{DBLP:conf/acl/QianHZDQCZZL0024} and CLAMBER \cite{DBLP:conf/acl/ZhangQDHLLJLC24}, with the hierarchical attributes in IN3 being crucial for assessing our model's prioritization mechanism.
To address the lack of targeted evaluation for visuospatial ambiguity, we introduce VRA-Ego (Visual and Referential Ambiguity in Egocentric view), a new benchmark of 1000 samples captured with modern AR glasses (e.g., Ray-Ban Meta \shortcite{Meta}, RayNeo X2/X3 Pro \shortcite{rayneo}). VRA-Ego is composed of two purpose-built subsets:
\begin{itemize}
\item Visual Ambiguity Set (500 images): This subset features intentionally flawed visual data (e.g., blur, poor framing). The ground truth consists of the precise corrective guidance needed to resolve the issue.
\item Referential Ambiguity Set (500 samples): This subset focuses on grounding deictic gestures, containing pointing actions, ambiguous queries, and their corresponding annotated answers.
\end{itemize}
The entire dataset is meticulously curated for diversity across scenes, lighting, object distances, and hand-object configurations to ensure real-world robustness.

\subsubsection{Evaluation Metrics.}

To evaluate our framework, we adopt a multi-faceted approach that assesses both component-level success and the semantic quality of the final output. For textual disambiguation, we first measure initial Vagueness Judgement Accuracy and dialogue efficiency through Average Conversation Rounds. The core performance is then captured by a Missing Details Recover Rate, calculated via a novel automated pipeline that simulates interaction to test the recovery of critical attributes. This principle of evaluating nuanced recovery extends to the vision module, where we assess not only the initial Target Identification Accuracy but also the quality of corrective feedback using a Strict and a Loose Recover Rate—the latter crediting partially correct suggestions (e.g., ``left" for ``top-left"). Ultimately, for cross-modal clarification, we evaluate pipeline applicability with Pointing Success Accuracy and measure the final output quality with a Semantic Answer Recover Rate. Here, to transcend the limitations of simple string matching and assess true comprehension, we employ a LLM judge to score the semantic similarity between the generated answer and the ground truth.        % powerful

\begin{figure*}[!t]
    \centering
    \includegraphics[width=0.8\linewidth]{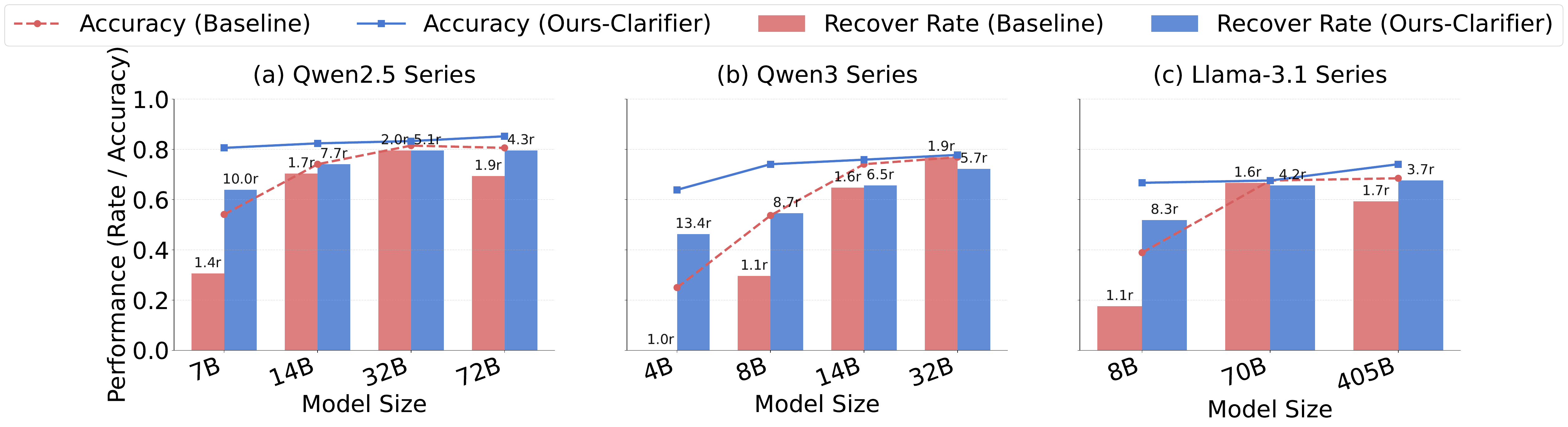}
    \caption{A comparative analysis of our proposed framework against the baseline across three open-source LLM families: (a) Qwen2.5, (b) Qwen3, and (c) Llama-3.1. The evaluation spans a range of model sizes to assess performance scalability. The primary performance metric, Recover Rate (representing the recovery of critical missing details), is shown using bar charts for both our method (blue) and the baseline (red). Additionally, the Accuracy (Baseline) is plotted as a dashed red line, while the Average Conversation Rounds (denoted by 'r' values) are annotated above each bar to measure dialogue efficiency.}
    \label{fig:aaai_multi_model_comparison}
\end{figure*}

\subsection{Comparison to Existing Methods}

\subsubsection{Choice of Baseline Model and Implementation Details.}

Our framework is designed as a zero-shot, external programmatic loop that enhances existing foundation models. To isolate and validate its benefits, we benchmark it against strong monolithic baselines across three core clarification tasks. Our method is evaluated by augmenting a diverse set of foundation models (Qwen, Llama, GPT-4o, Gemini, etc.), creating enhanced versions we denote with a -Clarifier suffix. The experimental comparisons are as follows:
\begin{itemize}
    \item Textual Disambiguation: We compare our iterative, multi-step clarification process against a standard monolithic prompt baseline. The baseline performs all reasoning steps (analysis, question generation) within a single, complex prompt, whereas our method guides the LLM through a sequence of targeted queries.
    \item Visual Object Grounding: The baseline VLM receives only the user query and image. In contrast, the VLM-Clarifier is augmented with our module that first invokes an open-set object detector and then performs a deterministic quality assessment on the resulting bounding box before proceeding.
    \item Cross-Modal Referential Resolution: The baseline VLM processes the raw image and query directly. The corresponding VLM-Clarifier is enhanced by our multi-stage pipeline that integrates depth estimation, hand segmentation, and 3D ray-casting to robustly interpret pointing gestures before feeding the identified target to the VLM.
\end{itemize}
For additional implementation details and baseline analyses (GPT-4 \cite{DBLP:journals/corr/abs-2303-08774}, Mistral variants \cite{DBLP:journals/corr/abs-2310-06825, DBLP:conf/acl/QianHZDQCZZL0024}), please refer to the Appendix.
% Supplementary Materials
% Appendix

\subsubsection{Quantitative Analysis.}

% Our quantitative analysis demonstrates the efficacy of our framework across textual, visual, and cross-modal clarification tasks. 
Our quantitative evaluation shows that our modular and programmatic framework 
achieves significant performance improvements on textual, visual, and 
cross-modal clarification tasks. 
% We consistently observe that our modular, programmatic approach provides significant performance gains by either scaffolding the reasoning of smaller models or by integrating deterministic algorithms for tasks where Vision-Language Models (VLMs) are inherently weak.
These improvements are driven by the framework's ability to either scaffold the reasoning of smaller models or to incorporate deterministic algorithms for tasks where Vision-Language Models (VLMs) are known to perform poorly.

\begin{table}[!t]
\centering
\resizebox{\columnwidth}{!}{%
\begin{tabular}{cccc}
\hline
\multirow{2}{*}{Model} & \multirow{2}{*}{\begin{tabular}[c]{@{}c@{}}Parameter\\ Size\end{tabular}} & \multicolumn{2}{c}{Accuracy (\%) $\uparrow$} \\ \cline{3-4} 
                           &      & w/o. Clarifier & w. Clarifier \\ \hline
\multirow{4}{*}{\shortstack{Qwen2.5\\\cite{DBLP:journals/corr/abs-2412-15115}}}   & 7B   & 24.4 & 53.0 (+ \textbf{28.6}) \\
                           & 14B  & 56.0 & 61.8 (+ \textbf{5.8}) \\
                           & 32B  & 56.0 & 66.0 (+ \textbf{10.0}) \\
                           & 72B  & 54.2 & 65.4 (+ \textbf{11.2}) \\ \midrule
\multirow{3}{*}{\shortstack{LLama-3.1\\\cite{DBLP:journals/corr/abs-2407-21783}}} & 8B   & 25.9 & 52.9 (+ \textbf{27}) \\
                           & 70B  & 53.5 & 59.3 (+ \textbf{5.8})\\
                           & 405B & 54.9 & 60.1 (+ \textbf{5.2})\\ \bottomrule
\end{tabular}%
}
\caption{Vagueness Judgement Accuracy on the CLAMBER benchmark. Our iterative Clarifier framework substantially boosts performance over single-prompt baselines. The framework yields an approximate 30\% increase for smaller models like Llama-3.1-8B and Qwen2.5-7B, elevating them to be competitive with much larger counterparts. This demonstrates that our structured, step-by-step approach acts as a critical reasoning scaffold, enabling smaller models to handle complex clarification tasks reliably.}
\label{tab:my-table-3}
\end{table}

\begin{table*}[!t]
\centering
\resizebox{\textwidth}{!}{%
\begin{tabular}{llllcl}
\toprule
\multicolumn{1}{c}{\multirow{3}{*}{Model}} &
  \multicolumn{3}{c}{(a) Vision-based Clarification} &
  \multicolumn{2}{c}{(b) Cross-Modal Pointing} \\ %\cline{2-6} 
  \cmidrule(lr){2-4}\cmidrule(lr){5-6}
    
\multicolumn{1}{c}{} &
  \multicolumn{1}{c}{\multirow{2}{*}{Accuracy (\%) $\uparrow$}} &
  \multicolumn{2}{c}{Recover Rate (\%) $\uparrow$} &
  \multirow{2}{*}{Accuracy (\%) $\uparrow$} &
  \multirow{2}{*}{Recover Rate (\%) $\uparrow$} \\ \cmidrule(l{0.5em}r{0.5em}){3-4}
\multicolumn{1}{c}{}    & \multicolumn{1}{c}{} & Strict          & Loose           &      &                \\ \midrule
gemini-2.5-pro \cite{google2025gemini25pro}          & 91.8                 & 46.2            & 60.2            & -    & 67.4           \\
gemini-2.5-pro-Clarifier & 95.4 (+ \textbf{3.6})       & 64.6 (+ \textbf{18.4}) & 75.8 (+ \textbf{15.6}) & 95.2 & 72.6 (+ \textbf{5.2}) \\
GPT-4o \cite{DBLP:journals/corr/abs-2410-21276}                   & 90.0                 & 40.6            & 57.0              & -    & 65.2           \\
GPT-4o-Clarifier         & 92.4 (+ \textbf{2.4})       & 61.4 (+ \textbf{20.8}) & 73.6 (+ \textbf{16.6}) & 94.4 & 69.0 (+ \textbf{3.8}) \\
Qwen2.5-VL \cite{DBLP:journals/corr/abs-2502-13923}               & 86.6                 & 35.4            & 53.2            & -    & 57.8           \\
Qwen2.5-VL-Clarifier     & 91.2 (+ \textbf{4.6})       & 47.4 (+ \textbf{12.0}) & 70.0 (+ \textbf{16.8}) & 92.4 & 64.4 (+ \textbf{6.6}) \\
llava-v1.6 \cite{li2024llavanext-strong}              & 84.8                 & 36.6            & 41.6            & -    & 53.6           \\
llava-v1.6-Clarifier     & 89.8 (+ \textbf{5.0})       & 53.2 (+ \textbf{16.6}) & 52.8 (+ \textbf{11.2}) & 91.6 & 58.0 (+ \textbf{4.4}) \\
InternVL 3.0 \cite{zhu2025internvl3}             & 83.2                 & 35.6            & 59.6            & -    & 51.6           \\
InternVL 3.0-Clarifier   & 88.6 (+ \textbf{5.4})       & 50.2 (+ \textbf{14.6}) & 70.2 (+ \textbf{10.6}) & 93.6 & 57.8 (+ \textbf{6.2}) \\
Llama 3.2 \cite{DBLP:journals/corr/abs-2407-21783}               & 82.6                 & 35.6            & 51.6            & -    & 50.2           \\
Llama 3.2-Clarifier      & 85.4 (+ \textbf{2.8})       & 51.2 (+ \textbf{15.6}) & 64.6 (+ \textbf{13.0}) & 91.2 & 53.8 (+ \textbf{3.6}) \\
MiniCPM-V \cite{DBLP:journals/corr/abs-2408-01800}                 & 81.6                 & 31.6            & 48.8            & -    & 47.0           \\
MiniCPM-V-Clarifier        & 86.2 (+ \textbf{4.6})       & 43.6 (+ \textbf{12.0}) & 63.4 (+ \textbf{14.6}) & 88.2 & 52.0 (+ \textbf{5.0}) \\
Molmo \cite{DBLP:conf/cvpr/DeitkeC0T0PSMLS25}                    & 80.0                 & 37.6            & 48.6            & -    & 48.4           \\
Molmo-Clarifier          & 84.0 (+ \textbf{4.0})       & 52.8 (+ \textbf{15.2}) & 62.2 (+ \textbf{13.6}) & 87.2 & 51.6 (+ \textbf{3.2}) \\ \bottomrule
\end{tabular}
}
\caption{Performance on Vision-based and Cross-Modal Pointing Clarification. The table compares baseline VLMs against the same models enhanced by our framework (suffixed with -Clarifier) on two egocentric tasks: (a) Evaluates corrective guidance for imperfect visual input. Here, Accuracy measures initial object identification, while Recover Rate (Strict/Loose) assesses the quality of the generated guidance. (b) Tests grounding of deictic queries via 3D pointing. Here, Accuracy (N/A for baselines) is our framework's success in detecting pointing intent, and Recover Rate measures the final answer's semantic accuracy.} % Our framework leverages specialized modules (geometric analysis, 3D ray-casting), whereas baselines rely on end-to-end inference.
\label{tab:my-table-2}
\end{table*}

The benefit of our iterative framework is most pronounced for smaller language models. As shown in Figure \ref{fig:aaai_multi_model_comparison}, our method improves the Recover Rate for smaller models like Qwen3-4B and Qwen2.5-7B by 43\% and 32\% respectively. This is because our step-by-step process decomposes a complex task into manageable sub-problems, which often overwhelms the single-turn capabilities of these models. While still beneficial, the performance margin narrows for larger models (e.g., Llama-3.1-405B) that are already proficient with complex, monolithic prompts. This trend is corroborated on the CLAMBER dataset (Table \ref{tab:my-table-3}), where our framework elevates the vagueness judgment accuracy of smaller models by nearly 30\%, making them competitive with much larger counterparts.

Our framework also shows strong performance on visuospatial tasks that challenge the geometric reasoning capabilities of end-to-end VLMs. 
For example, baseline models can identify visual referents but fail to provide precise spatial corrective feedback, achieving low Strict Recover Rates for directional guidance (31.5\%--46.2\%) (Table~\ref{tab:my-table-2}). 
To overcome this, our \mbox{-Clarifier} module performs deterministic geometric analysis on the object's bounding box. This targeted approach improves the Strict Recover Rate by a large margin of 11.9\% to 20.7\%.

Similarly, monolithic VLMs struggle to interpret pointing gestures from raw images, with the top baseline reaching a Recover Rate of only 67.3\% (Table~\ref{tab:my-table-2}). 
Our framework addresses this limitation by employing a two-step process: it first detects the pointing intent with high accuracy (87.2\%--95.1\%) before engaging a specialized 3D ray-casting pipeline. 
This explicit geometric modeling leads to consistent gains in task success, improving the Recover Rate by 3.1\% to 6.6\% for all evaluated VLMs. 
For both visuospatial tasks, this decompositional strategy allows our framework to surpass the performance of purely learned systems, resulting in a more robust solution.

\subsubsection{Qualitative Results.}

Our qualitative analysis substantiates the quantitative results, demonstrating the framework's practical robustness and efficiency. The core of our approach is an iterative, step-by-step process that systematically reduces ambiguity across modalities.

In textual dialogues, this manifests as significantly lower inference latency. By using targeted, minimal prompts each turn rather than a single monolithic one, complex disambiguation is resolved efficiently. This structured methodology proves equally effective in vision-centric tasks (Figure \ref{fig:vision_clarification_overview}). Our framework reliably analyzes complex scenes like chessboards or fine-print menus—scenarios where monolithic baselines often fail due to unresolved visual ambiguity.

The framework's advantage is most pronounced in resolving cross-modal pointing in cluttered environments (Figure \ref{fig:pointing}). While baselines are often distracted by salient but incorrect objects, our method robustly grounds the user's deictic reference. It first reconstructs the 3D pointing ray, intersects it with the scene using the depth map, and then generates a context-aware crop that isolates the target while preserving the vital hand-object relationship. This focused input enables the VLM to succeed where it would otherwise fail. We note that the primary failure mode occurs when monocular depth estimation is inaccurate for thin or reflective surfaces, causing the ray to pass through the intended object. A detailed gallery of qualitative comparisons is in the Appendix.       % available

\subsection{Ablation Studies}

We conducted ablation studies to validate our key design choices and assess the sensitivity of our modular framework.

\begin{itemize}
    \item \textbf{Object Detector Robustness:} 
    We evaluated the framework's sensitivity to the choice of object detector. While substituting our default model with Florence-2 \cite{DBLP:conf/cvpr/0004WXDHL00Y24} caused a small decrease in performance (~1\% Accuracy, ~3\% Recover Rate), other models like YOLOE \cite{DBLP:journals/corr/abs-2503-07465} and YOLO-World \cite{DBLP:conf/cvpr/ChengSG0WS24} showed slightly larger drops (~2\% Accuracy, ~5\% Recover Rate). Nevertheless, the performance with all alternative detectors remained substantially higher than the baseline without our clarifier. This demonstrates that our core contribution provides a consistent benefit, independent of the underlying detection model.

    \item \textbf{Importance of Specialized Fingertip Detection:} To measure the impact of our custom fingertip detector, we substituted it with the standard MediaPipe library \cite{DBLP:journals/corr/abs-1906-08172}. 
    This led to a significant decrease in Pointing Success Accuracy (15\%). This is because MediaPipe is not robust to the challenging hand views—often occluded or low-resolution in egocentric data, which confirms the need for our tailored approach.
    
    \item \textbf{Effective VLM Input Grounding:} Finally, as detailed in the Appendix, our context-aware cropping strategy for grounding VLM input proved more effective than alternative methods like point-based segmentation or using the full image \cite{DBLP:journals/corr/abs-2505-07062}, further justifying our specific design.
\end{itemize}

Collectively, these results highlight a key finding: while our framework provides a robust scaffold that is not overly sensitive to the choice of a general object detector, its peak performance critically relies on components specifically tailored to the challenges of egocentric vision, such as our specialized fingertip detector.

% \section{Discussion and Conclusion}     % 6.75 - 7 pages
\section{Conclusion}

In this paper, we introduced a modular framework for resolving multimodal user intent ambiguity in egocentric vision. Our approach emulates a Chain-of-Thought process, decomposing complex and ambiguous queries into a sequence of simpler, verifiable sub-problems. These sub-problems are solved by specialized modules, including modules for textual analysis, visual quality assessment using a hybrid of a Vision-Language Model (VLM) and traditional algorithms, and 3D gesture grounding via ray-casting. A key result of our work is that this structured reasoning process significantly improves the performance of smaller language models (e.g., 7B models) on text intent disambiguation, making our approach practical for deployment on resource-constrained platforms, such as AR glasses. Our work shows that hybrid architectures, which combine the reasoning capabilities of large models with the precision of deterministic algorithms, present a promising direction for building more capable and reliable embodied AI. 
Future work will improve conversational efficiency and extend our approach to physically embodied agents that actively seek clarification.

\section{Acknowledgments}

This work was supported by the Shenzhen Science and Technology Program (Grant No. ZDSYS20220323112000-001).

% \clearpage

\bibliography{aaai2026}

\begin{thebibliography}{10}
\providecommand{\natexlab}[1]{#1}

\bibitem[{Chen et~al.(2025)Chen, Wang, Jiang, and Nakashima}]{DBLP:conf/aaai/ChenWJN25}
Chen, J.; Wang, B.; Jiang, Z.; and Nakashima, Y. 2025.
\newblock Putting People in LLMs' Shoes: Generating Better Answers via Question Rewriter.
\newblock In \emph{AAAI, Sponsored by the Association for the Advancement of Artificial Intelligence, Philadelphia, PA, {USA}}, 23577--23585. {AAAI} Press.

\bibitem[{Grauman et~al.(2025)Grauman, Westbury, Byrne et~al.}]{DBLP:journals/pami/GraumanWBCCFGHJKLLMNRRR25}
Grauman, K.; Westbury, A.; Byrne, E.; et~al. 2025.
\newblock Ego4D: Around the World in 3,600 Hours of Egocentric Video.
\newblock \emph{{IEEE} Trans. Pattern Anal. Mach. Intell.}, 47(11): 9468--9509.

\bibitem[{Guo, Wu et~al.(2025)}]{DBLP:journals/corr/abs-2505-07062}
Guo, D.; Wu, F.; et~al. 2025.
\newblock Seed1.5-VL Technical Report.
\newblock \emph{CoRR}, abs/2505.07062.

\bibitem[{Kurita, Katsura, and Onami(2023)}]{DBLP:conf/iccv/KuritaKO23}
Kurita, S.; Katsura, N.; and Onami, E. 2023.
\newblock RefEgo: Referring Expression Comprehension Dataset from First-Person Perception of Ego4D.
\newblock In \emph{{IEEE/CVF} International Conference on Computer Vision, {ICCV} 2023, Paris, France, October 1-6, 2023}, 15168--15178. {IEEE}.

\bibitem[{Li et~al.(2025)Li, Liu, Wu et~al.}]{DBLP:journals/corr/abs-2504-11277}
Li, G.; Liu, W.; Wu, Y.; et~al. 2025.
\newblock From Misleading Queries to Accurate Answers: {A} Three-Stage Fine-Tuning Method for LLMs.
\newblock \emph{CoRR}, abs/2504.11277.

\bibitem[{Nakamura et~al.(2023)Nakamura, Kawanishi, Nobuhara et~al.}]{DBLP:conf/iccv/NakamuraKNN23}
Nakamura, S.; Kawanishi, Y.; Nobuhara, S.; et~al. 2023.
\newblock DeePoint: Visual Pointing Recognition and Direction Estimation.
\newblock In \emph{{IEEE/CVF} International Conference on Computer Vision, {ICCV}, Paris, France}, 20520--20530. {IEEE}.

\bibitem[{Qian et~al.(2024)}]{DBLP:conf/acl/QianHZDQCZZL0024}
Qian, C.; et~al. 2024.
\newblock Tell Me More! Towards Implicit User Intention Understanding of Language Model Driven Agents.
\newblock In \emph{Proceedings of the 62nd Annual Meeting of the Association for Computational Linguistics}, 1088--1113. ACL.

\bibitem[{Ramrakhya et~al.(2025)Ramrakhya, Chang, Puig et~al.}]{DBLP:journals/corr/abs-2504-00907}
Ramrakhya, R.; Chang, M.; Puig, X.; et~al. 2025.
\newblock Grounding Multimodal LLMs to Embodied Agents that Ask for Help with Reinforcement Learning.
\newblock \emph{CoRR}, abs/2504.00907.

\bibitem[{Sarch et~al.(2025)Sarch, Kumaravel, Ravi et~al.}]{DBLP:conf/acl/SarchKRVW25}
Sarch, G.~H.; Kumaravel, B.~T.; Ravi, S.; et~al. 2025.
\newblock Grounding Task Assistance with Multimodal Cues from a Single Demonstration.
\newblock In \emph{Findings of the Association for Computational Linguistics, {ACL} 2025, Vienna, Austria, July 27 - August 1, 2025}, 12807--12833. Association for Computational Linguistics.

\bibitem[{Wang et~al.(2024)Wang, Zhang, Zohar, and Yeung{-}Levy}]{DBLP:conf/eccv/WangZZY24}
Wang, X.; Zhang, Y.; Zohar, O.; and Yeung{-}Levy, S. 2024.
\newblock VideoAgent: Long-Form Video Understanding with Large Language Model as Agent.
\newblock In \emph{Computer Vision - {ECCV} 2024 - 18th European Conference, Milan, Italy, September 29-October 4, 2024, Proceedings, Part {LXXX}}, volume 15138 of \emph{Lecture Notes in Computer Science}, 58--76. Springer.

\end{thebibliography}

\section{Appendix}

This document provides supplementary material for our AAAI 2026 submission, ``Plug-and-Play Clarifier: A Zero-Shot Multimodal Framework for Egocentric Intent Disambiguation.'' Here, we include extended methodological details, additional experimental results, in-depth analyses, and broader discussions that were omitted from the main paper due to space constraints. This material is intended to offer greater insight into our design choices, the robustness of our framework, and avenues for future research. 
For more details, please refer to the core code of our project.
% And we will make the dataset, code, and demo publicly available upon acceptance of the paper.

\section{Detailed Cross-Modal Referential Clarification}

A fundamental challenge in egocentric interaction is the robust resolution of referential ambiguity, where a user's verbal query containing deictic terms (e.g., ``this'', ``that'') must be precisely grounded to a specific physical object indicated by an accompanying pointing gesture. This necessitates effective visual grounding, which remains a non-trivial task, particularly in dynamic, unconstrained real-world environments.

\subsubsection{Object Identification and Context-Aware Cropping for VLM Input}

As mentioned in the main paper, our pipeline accurately identifies the focal point of attention via 3D ray-casting, yielding a precise intersection point. Upon localizing this focus, the subsequent critical challenge is to appropriately delineate the visual entity at this location for effective processing by a Vision-Language Model. Early explorations into direct grounding methods, such as point-prompted segmentation (e.g., using SAM) or open-vocabulary object detection, proved brittle in egocentric settings. These approaches were particularly susceptible to inaccuracies stemming from cluttered scenes, minor pointing ray discrepancies, or the inherent challenges of novel object recognition.

Naively submitting the full egocentric image to the VLM introduces considerable visual noise and irrelevant contextual information, which can dilute the model's focus. Conversely, a simplistic tight crop of only the putative object severs the crucial visual link between the user's deictic gesture (e.g., the pointing hand) and the target object, thereby undermining the referential context. These limitations critically underscored the necessity for a more sophisticated input strategy, leading us to develop our context-aware cropping approach.

\begin{figure}[ht!]
    \centering
    % Assuming visual-prompt.pdf contains all subfigures (a-h) arranged within it.
    % If they are separate files, you would use \subfloat for each and include them individually.
    \includegraphics[width=0.995\linewidth]{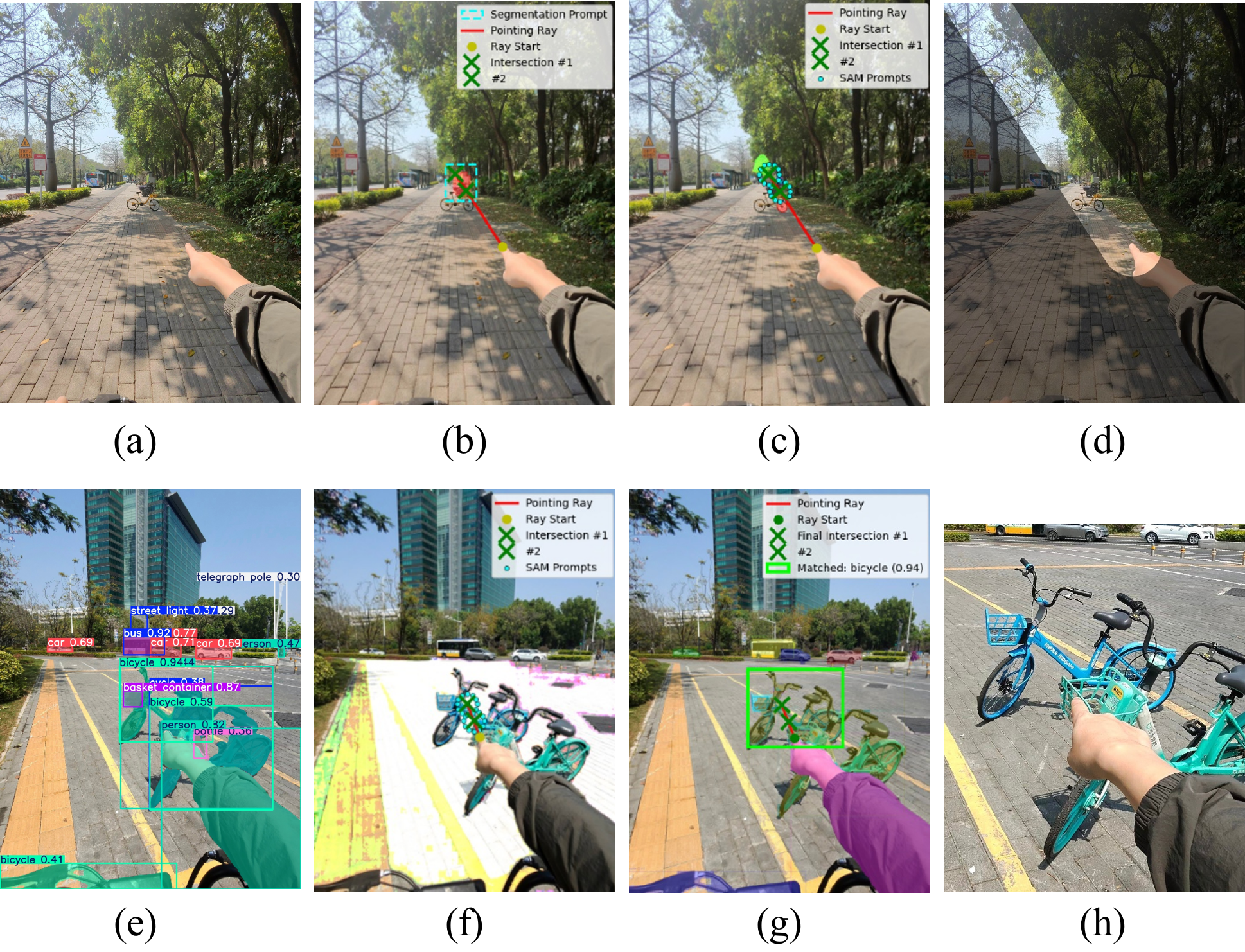}
    \caption{Various VLM input grounding strategies explored. (\textbf{a}) Original egocentric view. (\textbf{b-f}) Alternative rule-based and pre-processing strategies: (\textbf{b}) Intersection point connectivity graph, illustrating a graph-based segmentation approach. (\textbf{c}) Proximity-aware segmentation, where surrounding points are first gathered for mask generation. (\textbf{d}) Local attention derived from extended ray peripheral brightness, leveraging ambient light cues. (\textbf{e}) Recognition-first matching, where objects are identified before linking to the pointing gesture. (\textbf{f}) Segmentation-first matching, where scene segments are generated prior to association. (\textbf{g}) Bounding box rendered onto the original image. (\textbf{h}) \textbf{Our Adaptive Depth-Aware Contextual Cropping} method, which dynamically adjusts crop size based on object depth (further objects, larger crop) while ensuring the user's hand/arm is consistently included, preserving critical deictic context.}
    \label{fig:vlm_input_strategies}
\end{figure}

\begin{figure*}[!t]
    \centering
    \includegraphics[width=0.99\linewidth]{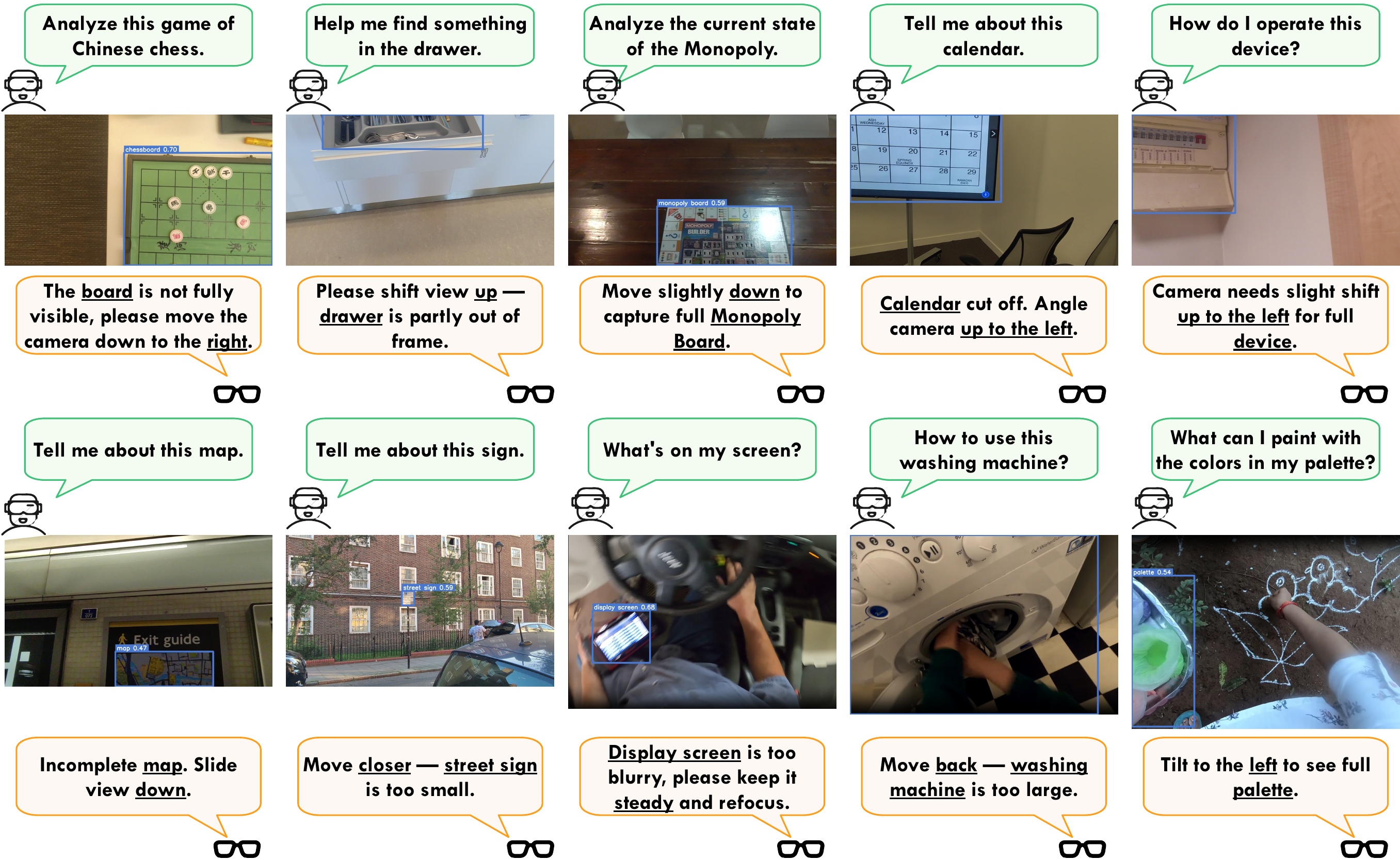}
    \caption{Demonstration of our framework's proactive visual clarification, a key differentiator from monolithic models. While baseline systems might fail or hallucinate when presented with flawed egocentric views, our method first diagnoses the input quality. It robustly identifies issues like occluded objects (\textit{Monopoly Board}), distance-related ambiguity (\textit{street sign}), and motion blur (\textit{display screen}), providing users with explicit instructions for correction. By resolving visual ambiguity at the source, our framework establishes a reliable foundation for subsequent high-level reasoning.}
    \label{fig:Appendix - 1}
\end{figure*}

\begin{figure*}[!t]
    \centering
    \includegraphics[width=0.99\linewidth]{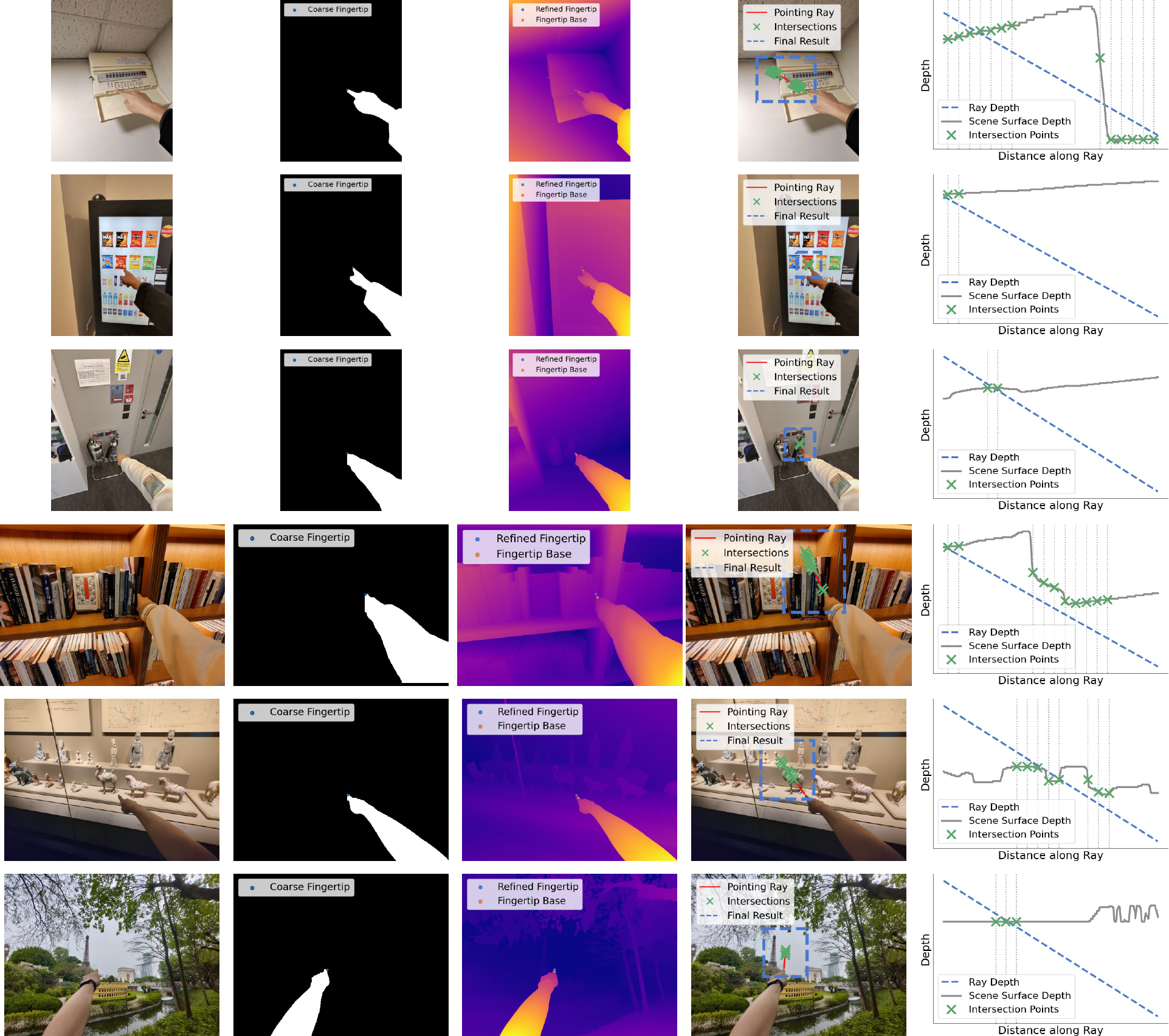}
    \caption{Visualization of our multi-stage pipeline for 3D pointing gesture interpretation across diverse egocentric scenes. Our method deterministically grounds deictic references through a sequence of geometric operations. From left to right: The process begins with the original egocentric view, followed by coarse hand segmentation to isolate the forearm. We then leverage a monocular depth estimate to refine the 3D locations of the fingertip and finger base. These points define a 3D pointing ray, which is cast into the scene. The final target is identified by analyzing the intersections between the ray and the scene's surface geometry. The {rightmost column} plots the ray's depth against the scene's surface depth, illustrating how our algorithm robustly identifies the correct intersection point, even in the presence of complex, non-planar surfaces (e.g., bookshelves) or distant targets. This geometric approach provides a robust alternative to end-to-end models, which often struggle with such spatial reasoning.}
    \label{fig:Appendix - 2}
\end{figure*}

\begin{figure*}[!t]
    \centering
    \includegraphics[width=0.99\linewidth]{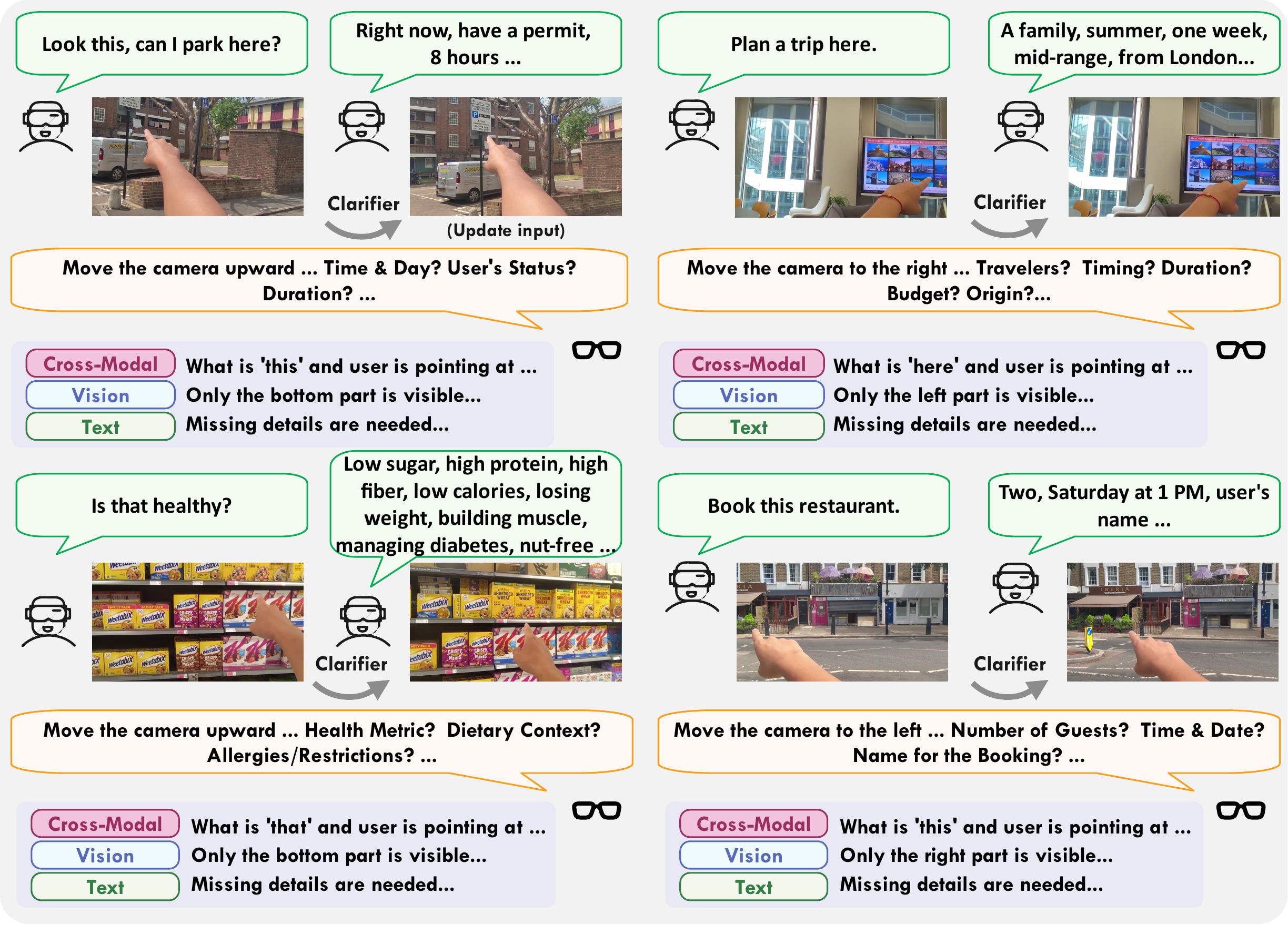}
    \caption{
    End-to-end examples of our framework resolving complex, co-occurring visual, textual, and cross-modal ambiguities in real-world egocentric scenarios. The area beneath the glasses icon illustrates the system's internal thought process, where it diagnoses different types of ambiguity. This diagnosis recognizes when the text is unclear because details are missing, when the vision is incomplete because an object is partially occluded, and how a cross-modal reference like the word "this" is connected to the user's pointing gesture. Based on this comprehensive analysis, the Clarifier generates a compound clarification request, prompting the user for both a physical camera adjustment and additional contextual information. This synergistic approach, which explicitly decomposes and addresses ambiguity across modalities, allows the framework to robustly handle initially underspecified tasks and transform them into solvable problems.
    }
    % \caption{End-to-end examples of our framework resolving complex, co-occurring visual, textual and and cross-modal (deictic pronoun) ambiguities in real-world egocentric scenarios. For each initial query (e.g., ``Can I park here?''), our system performs a multi-faceted diagnosis. It simultaneously identifies {visual ambiguity} (the parking sign is partially occluded) and {textual ambiguity} (the query lacks essential context like time of day or user's permit status). Consequently, the \mbox{-Clarifier} generates a \textit{compound clarification request}, prompting the user for both a physical camera adjustment and additional contextual information. This synergistic approach, which explicitly decomposes and addresses ambiguity across modalities, allows the framework to robustly handle initially underspecified tasks, transforming them into solvable problems.}
    \label{fig:Appendix - 3}
\end{figure*}

\subsubsection{Ablation Study of VLM Input Grounding Strategies}
As promised in the main paper, we conducted a comprehensive ablation study comparing our context-aware cropping against several alternative VLM input grounding strategies, visually summarized in Figure \ref{fig:vlm_input_strategies}.

\begin{itemize}
    \item \textbf{Full Image with Visual Cue} (Figure \ref{fig:vlm_input_strategies}a and \ref{fig:vlm_input_strategies}g): We provided the full egocentric image to the VLM, augmented with a visual cue (e.g., a green dot at the 2D projection, or a rendered bounding box \citeA{DBLP:journals/corr/abs-2505-07062}) to highlight the target. While performing marginally better than a completely naive full-image submission, this approach remained highly susceptible to distraction from other salient objects within the broad field of view, leading to a $\sim$4\% drop in Semantic Answer Recover Rate compared to our method. Furthermore, we critically observed that directly rendering bounding boxes (e.g., as depicted in Figure \ref{fig:vlm_input_strategies}g) onto the full image, intended to explicitly demarcate the target, consistently yielded sub-optimal performance compared to actual cropping. Our hypothesis, supported by empirical observations, is that not all large language models (LLMs) within the VLM architecture possess the inherent capability to effectively process or interpret such graphical overlays within the image plane, potentially perceiving them as noise or distractive elements rather than meaningful cues.

    \item \textbf{Rule-Based Grounding Strategies} (Figures \ref{fig:vlm_input_strategies}b-\ref{fig:vlm_input_strategies}f): Beyond simple point-prompted segmentation (Figure \ref{fig:vlm_input_strategies}c), we explored a suite of rule-based grounding approaches:
        \begin{itemize}
            \item \textbf{Intersection Connectivity Graph} (Figure \ref{fig:vlm_input_strategies}b): This method involved inferring object boundaries based on the local connectivity and cluster density of pixels immediately surrounding.
            \item \textbf{Proximity-Aware Segmentation} (Figure \ref{fig:vlm_input_strategies}c): Utilizing the 2D projection as a seed for a segmentation model (e.g., SAM) to generate a mask, then cropping the image to that mask. This was highly sensitive to the precision of the pointing ray and frequently failed on small or partially occluded objects, resulting in a $\sim$5\% performance drop.
            \item \textbf{Brightness-Weighted Local Context} (Figure \ref{fig:vlm_input_strategies}d): This approach extended the pointing ray and leveraged the average brightness of the surrounding area as a heuristic for local attention, assuming the target object would exhibit a distinct brightness profile from its immediate background.
            \item \textbf{Recognition-First Matching} (Figure \ref{fig:vlm_input_strategies}e): Here, a general object detection model was first employed to identify common objects across the scene, which were then post-hoc matched to the pointing ray.
            \item \textbf{Segmentation-First Matching} (Figure \ref{fig:vlm_input_strategies}f): This involved generating generic segmentation masks for all discernible objects in the scene, followed by associating the pointing ray with the most plausible mask.
        \end{itemize}
        While offering varying degrees of operational simplicity, we collectively found that these purely rule-based and pre-processing strategies (Figures \ref{fig:vlm_input_strategies}b-\ref{fig:vlm_input_strategies}f) were notably brittle and exhibited significant performance degradation when encountering less common or novel objects, or when tasked with complex Question-Answering (QA) and reasoning queries that demanded a deeper, semantic understanding of object context beyond simple geometric heuristics. Their inherent reliance on predefined rules severely limited their generalizability and robustness in unconstrained egocentric scenarios.

    \item \textbf{Our method} (Figure \ref{fig:vlm_input_strategies}h): Our proposed method adaptively determines the crop region based on the estimated depth of the target object. This ensures that objects further away are encompassed within a proportionally larger contextual window, providing sufficient surrounding information. Crucially, our strategy explicitly guarantees the consistent inclusion of the user's hand and a portion of their arm within the crop. This dynamic and contextually rich cropping strategy preserves the critical deictic gesture-object relationship while providing the VLM with a focused, yet semantically rich, input. This approach consistently outperformed all alternative strategies, yielding a substantial 3-5\% improvement in the final Semantic Answer Recover Rate over the next-best method. This robust empirical evidence unequivocally confirms that preserving the immediate hand-object interaction context is paramount for facilitating the VLM's robust reasoning capabilities in egocentric referential resolution.
\end{itemize}

\section{Extended Experimental Details}

\subsection{Novelty and Evaluation}

 Our core novelty lies in demonstrating that a hybrid, modular architecture—combining programmatic control with LLM reasoning—is a more robust and efficient paradigm for egocentric AI than monolithic models, especially for tasks requiring spatial precision where VLMs are known to be brittle. We created VRA-Ego because existing datasets do not target the specific problem of interactive, first-person referential disambiguation. For instance, DeePoint \citeA{DBLP:conf/iccv/NakamuraKNN23} (third-person), or RefEgo \citeA{DBLP:conf/iccv/KuritaKO23} (textual), differing from the direct "what is this?" pointing gestures in real-world assistance. Our work addresses this critical gap. Regarding generalization (e.g. Ego4D \citeA{DBLP:journals/pami/GraumanWBCCFGHJKLLMNRRR25}), our framework is a plug-in module that only activates when ambiguity is detected, otherwise defaulting to the base VLM's performance.

\subsection{VRA-Ego Dataset Construction and Setup}

Our ambiguity-oriented egocentric benchmark, VRA-Ego, was collected with current AI / AR glasses. The dataset covers pointing targets at distances from 0.2\,m up to 10\,m, so that the clarifier must handle both large, near-field objects and very small, far-field objects in the same pipeline. To make the cross-modal module robust to different user habits, we recorded both left- and right-hand pointing gestures, and we captured scenes in both portrait and landscape orientations. Each sample stores the RGB frame, the intended target object, and whether the raw capture was immediately answerable or required vision-based clarification. This information is later used by our automated evaluation pipeline to decide whether the controller should trigger the vision clarifier.

\subsection{Automated Evaluation Pipeline for Textual Disambiguation}
To comprehensively evaluate textual disambiguation and ensure scalable, reproducible assessment, we introduced a sophisticated automated evaluation pipeline. This three-stage process involves:
\begin{enumerate}
    \item \textbf{An Interaction Simulator} to generate complete dialogue logs by interacting with the model under test.
    \item \textbf{A Query Disentangler} to parse the agent's generated questions into atomic informational units.
    \item \textbf{A Semantic Matcher} that uses a powerful LLM judge (GPT-4o) to compare these atomic units against the ground-truth missing attributes from the benchmark, calculating the Recover Rate.
\end{enumerate}
This pipeline allows for nuanced evaluation beyond simple accuracy, capturing the model's ability to effectively extract specific, critical details.

\section{Additional Results and Analysis}
\label{sec:appendix-analysis}

\subsection{User Experience}
% failure cases
User experience and efficiency are critical. Qualitatively, our framework makes the assistant feel significantly more intelligent and responsive, as shown in our supplementary demo video. And the ablations are straightforward, reflecting our method's plug-and-play design.

\subsection{Additional Baseline Results for Textual Clarification}
In addition to the open-source model families presented in the main body of our work, we conducted a supplementary evaluation on the IN3 benchmark~\citeA{DBLP:conf/acl/QianHZDQCZZL0024}. This served to benchmark our \mbox{-Clarifier} framework against prominent proprietary models and other widely-used open-source variants. The results, detailed in Table \ref{tab:appendix-text-results}, corroborate our central thesis: our programmatic, iterative approach provides a critical reasoning scaffold, substantially enhancing the instruction-following capabilities of LLMs, especially for models that are less powerful or not extensively instruction-tuned.

Our initial step was to validate our experimental setup by replicating the results reported in the original IN3 paper (denoted by `*`). Our baseline implementations for GPT-4 and Mistral-Interact yielded performance metrics largely consistent with the reported figures, confirming the fidelity of our reproduction. However, a significant discrepancy emerged with Mistral-7B-v0.2. The monolithic baseline prompt, employed in the original work, failed to elicit valid, format-compliant responses from the model in our environment, resulting in a complete failure to produce measurable outcomes (indicated by `-`). We attribute this to the model's limited capacity to parse and execute complex, multi-part instructions embedded within a single prompt.

Furthermore, our investigation of the fine-tuned Mistral-Interact model revealed potential signs of overfitting within the original model's training regime. While the model performs well on the IN3 test set, its instruction-following capability proved brittle; minor variations to the prompt or query structure caused it to fail completely. This fragility prevented a robust evaluation of our \mbox{-Clarifier} on this specific model. We have contacted the original authors regarding this potential training artifact and details about baseline models performance but have not yet received a response.

In stark contrast, our framework demonstrates its efficacy most dramatically on models that struggle with monolithic prompting. For instance, Mistral-7B-v0.2, which failed outright under the baseline condition, becomes a competent performer when guided by our \mbox{-Clarifier}, achieving a 0.712 Accuracy. Even more strikingly, Mistral-7B-v0.3, whose baseline performance is near-random (0.213 Accuracy, 0 Recover Rate), is transformed into a highly effective system (0.806 Accuracy, 0.748 Recover Rate) with our framework. This substantial improvement—from complete failure to performance rivaling that of GPT-4—underscores the power of our structured, iterative decomposition in unlocking the latent capabilities of smaller language models.

\begin{table}[h]
\centering
\resizebox{\columnwidth}{!}{%
\begin{tabular}{lccc}
\toprule
\multicolumn{1}{c}{\textbf{Model}}  & \textbf{Accuracy} & \textbf{Recover Rate} & \textbf{Rounds} \\ \midrule
GPT-4*                     & 0.824    & 0.752        & 2.69   \\
GPT-4 (Baseline)           & 0.796    & 0.754        & 1.85   \\
\textbf{GPT-4-Clarifier}   & \textbf{0.815}    & \textbf{0.805}        & 3.71   \\ \midrule
Mistral-7B-v0.2*           & 0.491    & 0.684        & 1.62   \\
Mistral-7B-v0.2 (Baseline) & -        & -            & -      \\
\textbf{Mistral-7B-v0.2-Clarifier} & \textbf{0.712}    & 0.358        & 2.47   \\ \midrule
% Mistral-Interact*          & 0.852    & 0.723        & 4.15   \\
% Mistral-Interact (Baseline)& 0.806    & 0.740        & 3.51   \\
% Mistral-Interact-Clarifier & -        & -            & -      \\ \midrule
Mistral-7B-v0.3 (Baseline) & 0.213    & 0            & 1      \\
\textbf{Mistral-7B-v0.3-Clarifier} & \textbf{0.806}    & \textbf{0.748}        & 12.07  \\ \bottomrule
\end{tabular}%
}
\caption{Performance comparison for textual clarification on the IN3 benchmark. `*' denotes results from the original paper~\citeA{DBLP:conf/acl/QianHZDQCZZL0024}. `-' indicates the model failed to produce format-compliant outputs. Our \mbox{-Clarifier} framework not only enhances capable models like GPT-4 but also successfully scaffolds weaker models (e.g., Mistral-7B variants) that otherwise fail completely, elevating their performance to be competitive with much larger systems.}
\label{tab:appendix-text-results}
\end{table}

\subsection{Runtime and Latency Reporting}

To make the efficiency aspect of the clarifier explicit, we report the per-stage wall-clock latency on an RTX~4090 GPU. All numbers below are averaged over the full runs we used for the image-based (capture-quality) and cross-modal (pointing-based) clarifiers. We exclude one-time model loading and I/O bookkeeping.
In practice, the geometric stack (depth, pose, hand, fusion, ray) remains below 400\,ms per frame on this hardware, so the end-to-end latency is primarily determined by the choice and deployment of the LLM/VLM used for clarification.

% \paragraph{Image-based clarifier (capture-quality guidance).}
\begin{table}[h]
\centering
\begin{tabular}{l c}
\toprule
Stage & Avg. latency (ms) \\
\midrule
Image I/O              & 11.26 \\
Detection              & 185.52 \\
Feedback generation    & 4.47 \\
Visualization (optional) & 6.06 \\
\bottomrule
\end{tabular}
\caption{Per-stage latency for the image-based clarifier on RTX~4090. Detection is the dominant cost; feedback generation is negligible.}
\label{tab:image_latency}
\end{table}

% \paragraph{Cross-modal clarifier (pointing-based grounding).}
\begin{table}[h]
\centering
\begin{tabular}{l c}
\toprule
Stage & Avg. latency (ms) \\
\midrule
Depth estimation       & 268.62 \\
Pose detection         & 17.81 \\
Hand segmentation      & 39.19 \\
2D--3D pointing fusion & 43.83 \\
Ray intersection       & 5.50 \\
LLM processing         & 5684.59 \\
\bottomrule
\end{tabular}
\caption{Per-stage latency for the cross-modal clarifier on RTX~4090. The geometric part (depth, hand, ray) stays within a few hundred ms; the dominant cost is the LLM call. Development-time visualization was $\approx$3.7\,s and is excluded here because it is not part of the inference path.}
\label{tab:cross_modal_latency}
\end{table}

\noindent\textbf{Note on thresholds.} All ambiguity-related thresholds (e.g., blur score, minimum crop size, depth-aware expansion ratio) used in the above runs follow the default configuration shipped with our implementation. We did not heavily tune them for the reported numbers; a dedicated sensitivity study can further tighten these timings and trigger conditions.

\section{Broader Discussion and Future Work}

\subsection{Design Choices and Practicality}

Our modular design is a deliberate choice for real-time egocentric assistants. Unlike systems using video, large models, or complex "thinking" agents (e.g., VideoAgent \citeA{DBLP:conf/eccv/WangZZY24}), our approach prioritizes low latency. We argue that quick, iterative feedback more closely mimics natural human conversation, which is crucial for user experience. While gaze (e.g., in MICA \citeA{DBLP:conf/acl/SarchKRVW25}) could enhance clarity, its high cost, power consumption, and weight make it impractical for current consumer devices. Our framework provides a more readily deployable solution.

\subsection{Alternative Interaction Paradigms}
% During development, we considered several alternative approaches to ambiguity resolution that do not rely on a proactive clarification dialogue.
% \begin{itemize}
%     \item \textbf{Visual Ambiguity:} Unclear visual input (e.g., a blurry QR code) could be resolved if the user checks a camera preview on a screen, as is common with smartphones. However, current-generation AI glasses often lack such displays to maintain a lightweight form factor. This makes programmatic feedback essential.
%     \item \textbf{Textual Ambiguity:} An underspecified request could be clarified by having the user type a more detailed query. This is impractical for on-the-go, hands-free interaction, which is the primary use case for egocentric AI.
%     \item \textbf{Referential Ambiguity:} A user could avoid pointing by providing a very specific description of the target object (e.g., ``the third red book from the left on the top shelf''). However, this is unnatural and cognitively demanding compared to the seamless deictic gestures used in human conversation.
% \end{itemize}
% Our framework is designed specifically for the interaction paradigm of next-generation wearables and embodied agents, where proactive, low-friction clarification is paramount.

Our work introduces a modular, plug-and-play framework designed to proactively resolve multimodal intent ambiguity in egocentric interactions. While the quantitative results demonstrate significant performance gains, it is equally important to discuss our approach in the context of alternative design philosophies and the broader vision for human-AI collaboration.

One might argue that simpler, more direct solutions could address the ambiguities we tackle. For instance, visual ambiguity caused by poor framing or blurriness could ostensibly be resolved by providing the user with a real-time camera preview, much like one might adjust their phone during a video call. Similarly, referential ambiguity could be eliminated if the user provided a more explicit linguistic description (e.g., ``the second bottle from the left''), and underspecified textual intent could be clarified by requiring the user to input their complete request via a text box.

However, these solutions, while viable in constrained scenarios, fundamentally conflict with the vision of a seamless, conversational AI assistant that defines the next era of human-computer interaction. The ideal interaction paradigm is not a series of rigid, user-driven corrections but a fluid, natural dialogue. Imagine a phone conversation where nearly every statement required a pedantic clarification; the interaction would become intolerably cumbersome. Natural human dialogue is built on context and inference, and our goal is to imbue AI systems with this capability. Requiring explicit descriptions or manual camera adjustments imposes a cognitive load that disrupts this natural flow.

Furthermore, these manual-correction alternatives face significant practical and systemic drawbacks, especially for wearable devices like AR glasses. 
\begin{enumerate}
    \item \textbf{Limited Generalizability:} User-dependent clarification strategies do not readily generalize to autonomous agents, such as robots, which must interpret and act upon ambiguous instructions without constant human intervention. An effective disambiguation model must be part of the agent's core intelligence.
    \item \textbf{Power Consumption:} Constant camera previews or high-bandwidth video processing for user feedback are highly power-intensive, a critical bottleneck for all-day wearable devices.
    \item \textbf{User Experience and Hardware Limitations:} Real-time visual feedback can induce visual fatigue. Critically, most commercially available AI glasses are equipped with {fixed-focus lenses} and do not support autofocus, rendering any manual, user-driven adjustments for visual clarity largely impractical from the outset. Adding such capabilities would, in turn, increase the device's {weight} and complexity.
\end{enumerate}

Our Plug-and-Play Clarifier is designed precisely to navigate these challenges. By building a system that proactively identifies and resolves ambiguity through an intelligent, multimodal dialogue, our framework offloads the burden of clarification from the user to the agent. It paves the way for the kind of fluid, context-aware assistance that future egocentric AI promises, where interaction feels as natural as speaking with another person. This makes our approach not just an improvement, but a foundational necessity for the future of embodied, conversational AI.

\subsection{Extended Limitations and Future Work}
The main paper briefly touched on limitations. Here, we expand on those points and provide more context on future directions, drawing from our initial drafts.

Our approach introduces certain trade-offs. The iterative nature of our dialogue framework, while improving accuracy, inherently increases the number of conversational turns, which could affect user experience. Furthermore, as a pluggable module, the framework's overall efficacy is intrinsically tied to the speed and capability of the underlying foundation models. Our work also simplifies the multifaceted nature of real-world intent ambiguity. True ambiguity often stems from deep contextual dependencies or polysemy, where a query like ``How should I move?'' could refer to a chess move, a fashion catwalk, or a travel route, depending on a rich context that our current model does not fully leverage.

These limitations illuminate promising avenues for future research. To enhance efficiency, our framework could be augmented with proactive query rewriting techniques \citeA{DBLP:conf/aaai/ChenWJN25} to clarify user intent in a single step. To improve robustness, the system could incorporate mechanisms to detect and correct misleading or contradictory information within the user's query itself \citeA{DBLP:journals/corr/abs-2504-11277}, rather than only addressing what is missing. Ultimately, our framework provides a foundational stepping stone towards more sophisticated embodied agents. Future work can extend our principles to physical robots that must not only understand ambiguous commands but also actively navigate their environment to seek clarification and execute tasks \citeA{DBLP:journals/corr/abs-2504-00907}, moving us closer to creating truly intelligent agents capable of navigating the full spectrum of human communicative intent.

\bibliographystyleA{aaai2026}
\bibliographyA{aaai2026.bib}

\end{document}